\documentclass[conference]{IEEEtran}

\usepackage{silence}
\usepackage[utf8]{inputenc}
\usepackage[T1]{fontenc}

\usepackage{graphicx}
\usepackage{placeins} 

\usepackage{subfigure}
\usepackage[table]{xcolor}

\usepackage{amssymb,amsmath,amsthm}

\usepackage{booktabs}
\usepackage{algorithm}
\usepackage{algorithmic}
\usepackage[switch]{lineno}

\usepackage{array}
\usepackage{booktabs}
\usepackage{ragged2e}
\usepackage{makecell}
\usepackage{adjustbox}
\usepackage{threeparttable}
\usepackage{multicol,multirow}

\usepackage[hidelinks]{hyperref}
\usepackage{url}
\begin{document}

\title{Platelet enumeration in dense aggregates}

\author{\IEEEauthorblockN{Anonymous Authors}}

\author{
\IEEEauthorblockN{
H. Martin Gillis$^{1}$,
Yogeshwar Shendye$^{1}$,
Paul Hollensen$^{2}$, 
Alan Fine$^{2,3,4}$, and
Thomas Trappenberg$^{1*}$
}
\vspace{-0.2cm} \\
\IEEEauthorblockA{
$^1$Faculty of Computer Science, Dalhousie University, Halifax, NS, Canada
}
\IEEEauthorblockA{
$^2$Alentic Microscience Inc., Halifax, NS, Canada
}
\IEEEauthorblockA{
$^3$Department of Physiology and Biophysics, Dalhousie University, Halifax, NS, Canada
}
\IEEEauthorblockA{
$^4$School of Biomedical Engineering, Dalhousie University, Halifax, NS, Canada 
\vspace{0.2cm}
}
\{martin.gillis, yogeshwar.shendye, a.fine\}@dal.ca,
phollensen@alentic.com,
tt@cs.dal.ca
}
\maketitle

\renewcommand{\thefootnote}{\fnsymbol{footnote}}
\footnotetext[1]{Corresponding author:~\href{mailto:tt@cs.dal.ca}{tt@cs.dal.ca}}
\renewcommand{\thefootnote}{\arabic{footnote}}

% \linenumbers

% ABSTRACT
\begin{abstract}

Identifying and counting blood components such as red blood cells, various types of white blood cells, and platelets is a critical task for healthcare practitioners. 
Deep learning approaches, particularly convolutional neural networks (CNNs) using supervised learning strategies, have shown considerable success for such tasks. 
However, CNN based architectures such as U-Net, often struggles to accurately identify platelets due to their sizes and high variability of features. 
To address these challenges, researchers have commonly employed strategies such as class weighted loss functions, which have demonstrated some success. 
However, this does not address the more significant challenge of
platelet variability in size and tendency to form aggregates and 
associations with other blood components.
In this study, we explored an alternative approach by investigating the role of convolutional kernels in mitigating these issues.
We also assigned separate classes to singular platelets and platelet aggregates and performed semantic segmentation using various U-Net architectures for identifying platelets. 
We then evaluated and compared two common methods (pixel area method and connected component analysis) for counting platelets and proposed an alternative approach specialized for single platelets and platelet aggregates.
Our experiments provided results that showed significant improvements in the identification of platelets, highlighting the importance of optimizing convolutional operations and class designations. 
We show that the common practice of pixel area-based counting often over estimate platelet counts, whereas the proposed method presented in this work offers significant improvements.
We discuss in detail about these methods from segmentation masks. 

\end{abstract}

% INTRODUCTION
\vspace{0.5cm}
\section{Introduction}
\label{sect:introduction}
In computer vision, detecting even the smallest objects is crucial, especially in safety critical domains like medical image analysis and remote sensing~\cite{Chowdhury::2020a,Gurcan::2009a,Alentic.::2025a}.
For example, blood images contain extremely small objects such as platelets, which are blood components that help to stop bleeding. 
Accurate identification and detection of platelets plays an essential role in healthcare applications such as enabling reliable blood count estimations~\cite{WHO::2021a}.
These counts are critical for medical professionals to diagnose and monitor patient health effectively.
Healthcare professionals rely on centralized laboratory processing.
Due to high volume of patients, this often results in long delays for patient diagnosis.

\begin{figure}[!htb]
    \centering
    \includegraphics[width=0.42\textwidth]{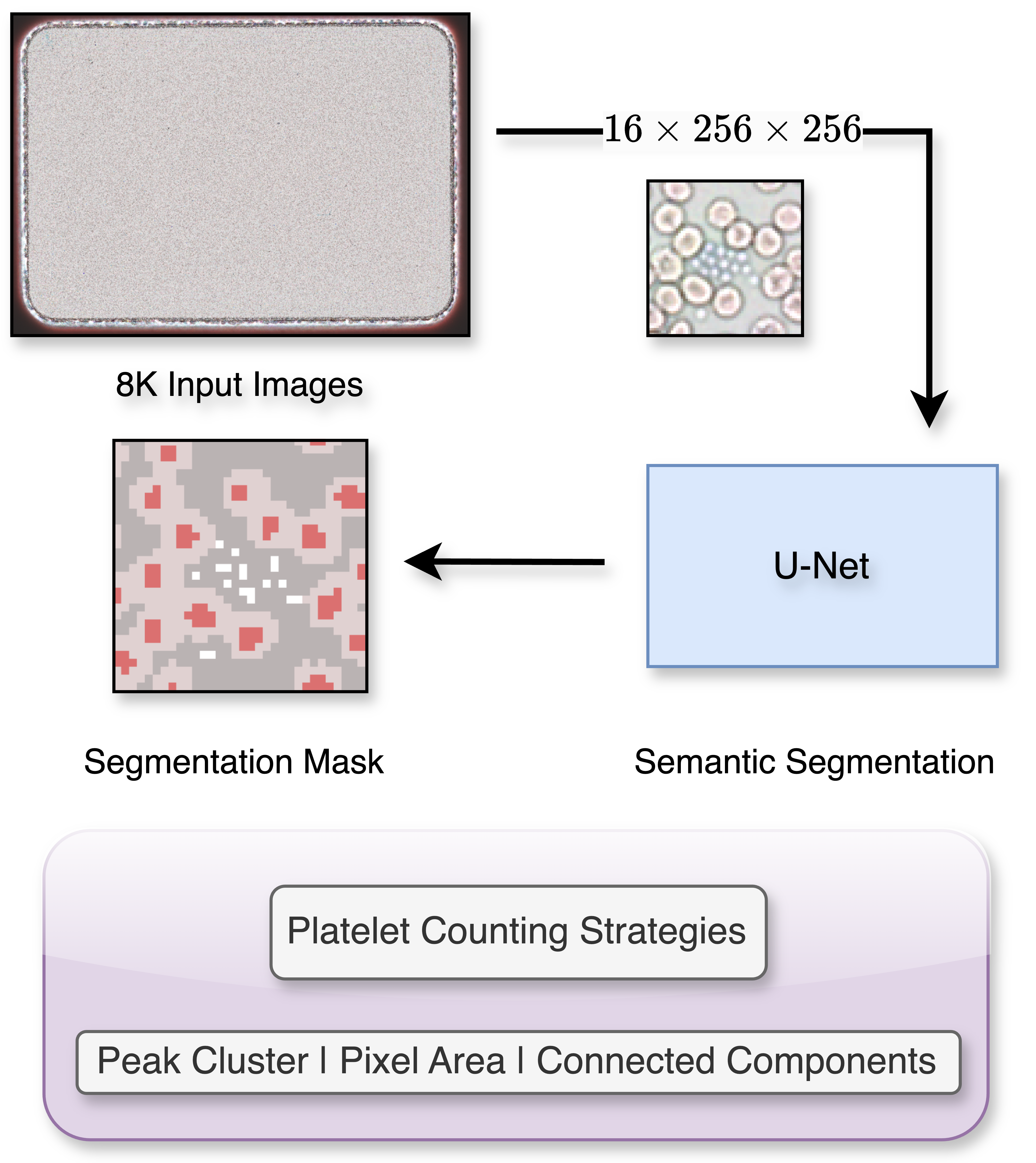}
    \caption{
    Overview of segmentation and platelet counting. 
    Images are cropped as $16 \times 256 \times 256$ images before using a U-Net network for semantic segmentation. 
    Platelet and platelet aggregate segmentation masks are then used to count platelets using three different approaches: 
    1) Peaks Cluster Method (proposed method); 
    2) Pixel Area Method; and 
    3) Connect Component Analysis.
    }
    \label{fig:overview_counting}
\end{figure}

\FloatBarrier

Deep learning network image processing techniques, such as semantic segmentation show promising results for automating such tasks.
U-Net is one such well-known architectures for semantic segmentation widely used for medical images~\cite{Ronneberger::2015a,Livermore::2022a}.
The U-Net architecture is symmetrical, mirroring the structure of an encoder-decoder design.
The encoder is a traditional convolutional network that applies successive convolutions and grouping layers to downsample the input image, thereby capturing increasingly abstract and complex features at smaller spatial resolutions.
This process results in a latent-space representation that encodes the essential characteristics of the input image.
The decoder conversely, uses upsampling layers and convolutional layers to incrementally restore the spatial dimensions of the feature maps, converting the latent representations back into the output image of the same size as the input.

Based on the segmentation masks generated by these networks, further computer vision techniques can be used to obtain counts for various blood cells and other important components~\cite{Singh::2024a}.
A common strategy for counting involves counting pixels for a given blood component and then normalizing by the average size of the component.
Since we are counting pixels in a 2D image array, this is often referred to as pixel area method (PAM).
The accuracy for this approach depends on factors like cell clustering (over-lapping cells), image quality, and variability in cell sizes and morphologies.
Another approach is connected component analysis (CCA) where labeled pixels are connected using 4-components (left-right-up-down) or 8-components (addition of diagonals).
A count here is defined where labeled pixels are continuously connected.
This method is also susceptible to over-lapping blood components, such as platelets and platelet aggregates.

In this work, we utilize the observation that platelets and their aggregates often have a very bright focal point which results with pixels corresponding to platelets with higher intensities.
We propose using a clustering algorithm to identify a cluster of similar pixels and then count the number of peaks with high intensities corresponding to a platelet.
A counting strategy we refer to as peak cluster method (PCM).
For cases which involve platelet aggregates, we anticipate this method to provide counts that are better than PAM an comparable to CCA. A summary of the proposed platelet counting strategies is illustrated in Fig.~\ref{fig:overview_counting}.
The main contributions of this research are summarized as:
\begin{itemize}
    \item We evaluated the role of convolution sizes used with U-Net networks for small object detection.
    \item We propose a peak cluster counting method for platelets and platelet aggregates.
    \item We validated the effectiveness of common counting techniques.
\end{itemize}

% METHODS
\section{Methods}
\label{sect:methods}
\subsection{Equipment and Data Collection}

Images were acquired using a lensless near-field microscope from Alentic Microscience, Inc.~\cite{Alentic.::2025a}.
This device provides high-resolution images (8K) from using multiple light-emitting diodes at different angles of projection.
Data collection is performed from a single drop of blood mixed with reagents.
After assembling the device and allowing time to establish a monolayer of blood, images are subsequently captured.
A process that takes less than 10 minutes to perform.
The images obtained after illumination have dimensions of $12 \times$ RGB images of dimensions $3 \times 3K \times 4K$, which were subsequently super-resolved to a dimension of $3 \times 6K \times 8K$.
Super-resolved images were then converted to 12-channels after which the addition of another set of RGB frames and an ultraviolet frame.
This provided input data with dimensions $16 \times 3K \times 4K$ and were used for the creation of datasets described in the next section~\cite{Alentic.::2025a}.

\subsection{Datasets and Preprocessing}

The images acquired from blood samples contain multiple occurrences of one or more of the following classes:

\begin{enumerate}
    \item background; 
    \item white blood cells (WBC); 
    \item platelets; 
    \item red blood cells (RBC, exterior); 
    \item red blood cells (RBC, interior); 
    \item beads; 
    \item artifacts; 
    \item debris; and 
    \item bubbles.
\end{enumerate}

\noindent
In the next sections, descriptions for the preparation of datasets, network configurations, hyperparameters, and training is provided for the U-Net semantic segmentation networks. 
The following sections will then described the methods used for counting platelets and platelets in the context of their aggregates.

\subsection{Network Implementations}
Conventional U-Net architectures has the number of output filters set to 64. 
The first layer of U-Net network takes the 16-channel input image and outputs 64-channels. 
The first downsampling layer increases the number of output channels from 64 to 128, second layer does the same from 128 to 256, third layer goes from 256 to 512 and the fourth one increases the number of output channels to 1024. 
To maintain the spatial features at higher resolutions learned by each of these downsampling layers, skip connections are used. 
From the bottleneck latent representation, upsampling layers increases the dimension and at the same time reducing the number of filters. 
Based on number of output filters of the first convolutional layers, network architectures are named accordingly. 
U-Net-16, U-Net-32, U-Net-64 and U-Net-128 corresponds to the U-Net architectures with 16, 32, 64 and 128 output channels for the first convolutional layers, respectively. 
In addition, experiments were performed with U-Net architectures where the down- and upsampling blocks were repeated only once (\textit{i.e.}, \texttt{convolution-batch normalization-activation}, referred to as a 1-stage down- and upsampling), which are named U-Net-64S.
All the models discussed in this section were trained and evaluated on the same training and validation dataset using the class weighted focal loss objective function.
Class weights were calculated as the square root of the inverse class frequency, which was subsequently normalized.
All models were trained over 40 epochs (batch size 16) using stochastic gradient decent with an initial learning rate of $1.0 \times 10^{-2}$ and a weight decay of 0.98 per epoch.

\subsection{Methods used for Counting Platelets}

Our first requirement to apply the proposed peak cluster method (PCM) is to identify and locate platelet aggregates.
Using the platelet aggregate segmentation masks from U-Net networks, we applied density-based spatial clustering

\begin{table*}[!htb]
\begin{adjustbox}{width=0.98\textwidth}
    \centering
    \begin{threeparttable}
        \caption{Performance Metrics for U-Net Semantic Segmentation Networks.\tnote{1}}
        \begin{tabular}{ l l l l l l l l }
            \toprule
            % headings
            \textbf{Network}
            & \textbf{Accuracy}
            & \textbf{F1-score}
            & \multicolumn{5}{c}{\textbf{F1-score}} \\
            \cmidrule(lr){5-8}
            & 
            & 
            & \textbf{Platelet} 
            & \textbf{WBC} 
            & \textbf{RBC (interior)} 
            & \textbf{RBC (exterior)} 
            & \textbf{Bead}
            \\
            \midrule
            U-Net-16 % 1 platelet class
            & $0.989 \pm 0.002$
            & $0.962 \pm 0.008$
            & $0.332 \pm 0.016$
            & $0.805 \pm 0.104$
            & $0.631 \pm 0.137$
            & $0.757 \pm 0.211$
            & $0.624 \pm 0.150$
            \\
            U-Net-32 % 1 platelet class
            & $0.989 \pm 0.002$
            & $0.962 \pm 0.008$
            & $0.304 \pm 0.152$
            & $0.821 \pm 0.079$
            & $0.527 \pm 0.222$
            & $0.721 \pm 0.218$
            & $0.543 \pm 0.177$
            \\
            U-Net-64 % 1 platelet class
            & $0.989 \pm 0.002$
            & $0.962 \pm 0.008$
            & $0.655 \pm 0.187$
            & $0.862 \pm 0.092$
            & $0.867 \pm 0.083$
            & $0.896 \pm 0.152$
            & $0.877 \pm 0.068$
            \\
            U-Net-128 % 1 platelet class
            & $0.989 \pm 0.002$
            & $0.962 \pm 0.008$
            & $0.745 \pm 0.043$
            & $0.939 \pm 0.017$
            & $0.900 \pm 0.015$
            & $0.949 \pm 0.007$
            & $0.880 \pm 0.039$
            \\
            \midrule
            U-Net-64S % 1 platelet class
            & $0.989 \pm 0.002$
            & $0.962 \pm 0.008$
            & $0.874 \pm 0.027$
            & $0.968 \pm 0.006$
            & $0.918 \pm 0.009$
            & $0.968 \pm 0.004$
            & $0.944 \pm 0.007$
            \\
            U-Net-64S2\tnote{2} % 2 platelet classes
            & $0.989 \pm 0.002$
            & $0.962 \pm 0.008$
            & $0.909 \pm 0.031$
            & $0.968 \pm 0.009$
            & $0.936 \pm 0.019$
            & $0.971 \pm 0.007$
            & $0.920 \pm 0.038$
            \\
            U-Net-64S2+\tnote{2,3} % 2 platelet classes (same as A3, trained on more labeled data)
            & $0.989 \pm 0.002$
            & $0.962 \pm 0.008$
            & $0.969 \pm 0.010$
            & $0.977 \pm 0.008$
            & $0.950 \pm 0.015$
            & $0.977 \pm 0.005$
            & $0.957 \pm 0.030$
            \\
            \midrule
            \bottomrule
        \end{tabular}
        $^1$ Metrics represents 10-fold cross-validation providing mean values along with their corresponding standard deviations. Values for accuracy and F1-scores represents all classes. Individual background, artifacts, debris, and bubbles omitted for clarity.
        $^2$ Models trained using 2 platelet classes: single platelet and platelet aggregates.
        $^3$ Trained with additional labels for platelet aggregates.
        \label{tab:f1_scores}
    \end{threeparttable}
\end{adjustbox}
\end{table*}

\FloatBarrier

\noindent
with noise (DBSCAN) algorithm~\cite{Ester::1996a} to access the coordinates of platelet aggregates as clusters. 
Since we are focusing on platelet aggregates, this algorithm enables us to access these aggregates individually as clusters and use their coordinates to obtain platelet counts for the PAM and PCM methods.
We configured the DBSCAN clustering algorithm with epsilon set to 1 and minimum samples set to 1. 
By setting epsilon to 1, we ensure that any two pixels within a distance of 1 unit would be considered neighbors. 
With minimum samples set to 1, even a single pixel would be classified as a platelet, as long as it
had at least one pixels (including itself) within this specified distance.
This configuration allowed us to form a large number of small clusters, capturing detailed groupings even in the case of sparse data.
For each cluster, we now have its position in the form of pixel coordinates within a mask. 
A bounding box is then calculated for the smallest enclosing boundary around each cluster.
The size of this bounding box was based on the largest of width or height for the cluster, ensuring consistency in extracted clusters across different samples. 
With the bounding region established, we next extract the corresponding platelet aggregate from the original image and the mask. 
To ensure we capture all relevant details, we extend the boundaries by adding a 5 pixel margin around the detected region. 
This extra padding helps to preserve contextual information that may be important for further analysis, preventing the loss of critical features near the edges. 
Once the cropped region is obtained, we analyze its intensity values to identify local maxima, which corresponds to the highest pixel values within the extracted area. 
By applying a filtering technique that compares each pixel to its neighboring pixels, we detect these high-intensity values.
To refine the platelet count selection, we apply a threshold value (0.9) that ensures only the most significant peaks are reported as a platelet count.
In the case of the pixel area method, rather than reporting platelet counts for threshold peak values, platelet counts are reported as number of pixels in the cluster normalized by the average size of a platelet (defined as 3 pixels).
Lastly, for connected component analysis, manual inspection of platelet segmentation masks were performed using 4-components to count distinct platelets.

% EXPERIMENTS
\section{Experiments}
\label{sect:experiments}
\subsection{Semantic Segmentation using U-Net Networks}

We performed 10-fold cross-validations for different sized U-Net architectures.
In Table~\ref{tab:f1_scores}, the results for using a U-Net-16 was 98.9\% with a F1-score of 96.2\%.
While this overall metric provides an acceptable results, upon closer inspection of specific blood components, the resulting F1-scores indicate otherwise.
In fact, platelets resulted with an F1-score of 33.2\%, which is not acceptable for any real application.
If we increase the size of the U-Net networks (\textit{i.e.}, U-Net-32, -64, and -128), the results did improve; however, they were still below any acceptable criteria for blood component analyses.
Despite that all networks resulted with high overall accuracies and F1-scores.
The above networks used the 2-stage down- and upsampling blocks (\textit{i.e.}, $2 \times \texttt{convolution-batch normalization-activation}$), which is commonly used for standard U-Net networks.
We hypothesize that since blood sample images contain small pixel-sized objects, we lose important spatial information that is captured with the residual skip connections of the network.
As a result, we opted to use only a 1-stage down- and upsampling approach to ameliorate this effect.
The network U-Net-64S shows that this made a significant improvement compared a 2-stage approach. 
In this case, we obtained a platelet F1-score of 87.4\%; although, the overall accuracy and F1-score were the same.
We next added a platelet aggregate class and trained the same network.
Here, U-Net-64S2 provided the same overall score for accuracy and F1-score but with improved platelet F1-score (90.9\%).
The U-Net-64S2+ network shows the results with additional labels used for platelet aggregates with a platelet F1-score of 96.9\%.
These results suggest that larger networks may not provide improved performance.
Rather, understanding the specific task and how the network obtain feature representations may play a more important role.
In the case of platelets and similar small objects in the context of a larger image, retaining spatial features at higher resolution may be a more important factor.

\begin{figure*}[!htb]
    \centering
    \subfigure[U-Net-64S]  
    {
        \includegraphics[width=0.48\textwidth]{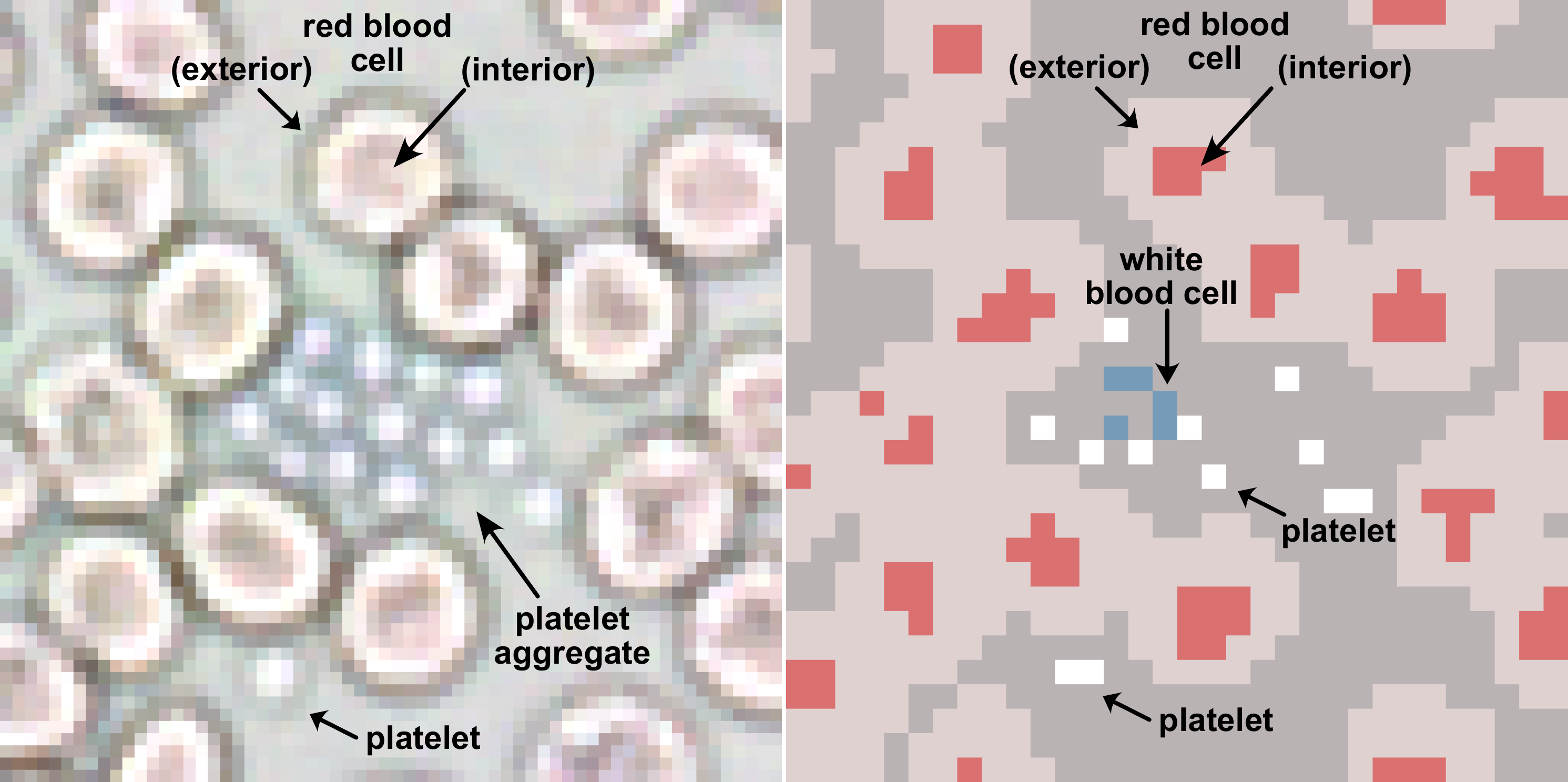}
        \label{fig:u64sl}
    }
    \subfigure[U-Net-64S2+] 
    {
        \includegraphics[width=0.48\textwidth]{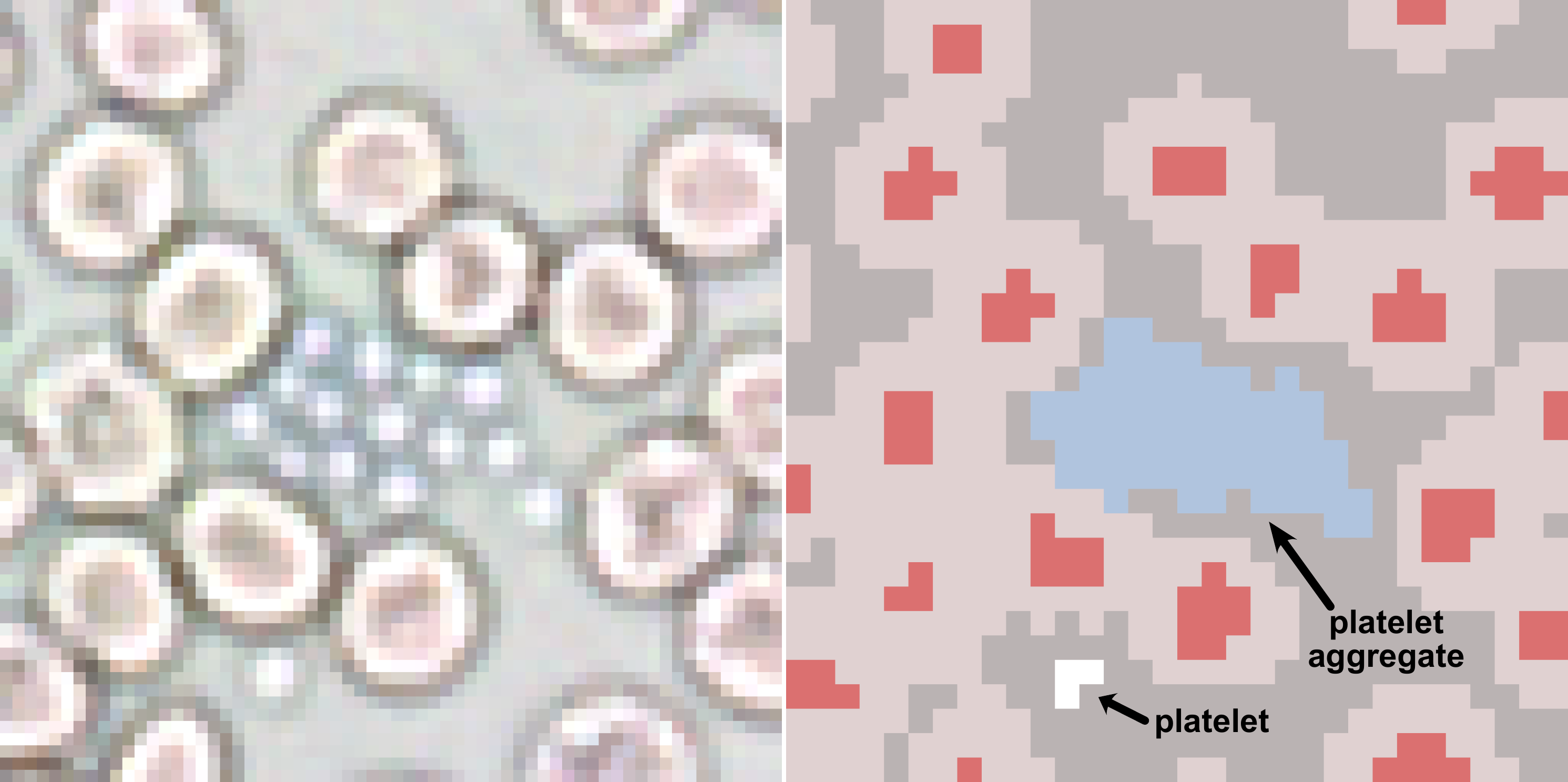}
        \label{fig:u64_2_classes}
    }
    \subfigure[U-Net-16] 
    {
        \includegraphics[width=0.48\textwidth]{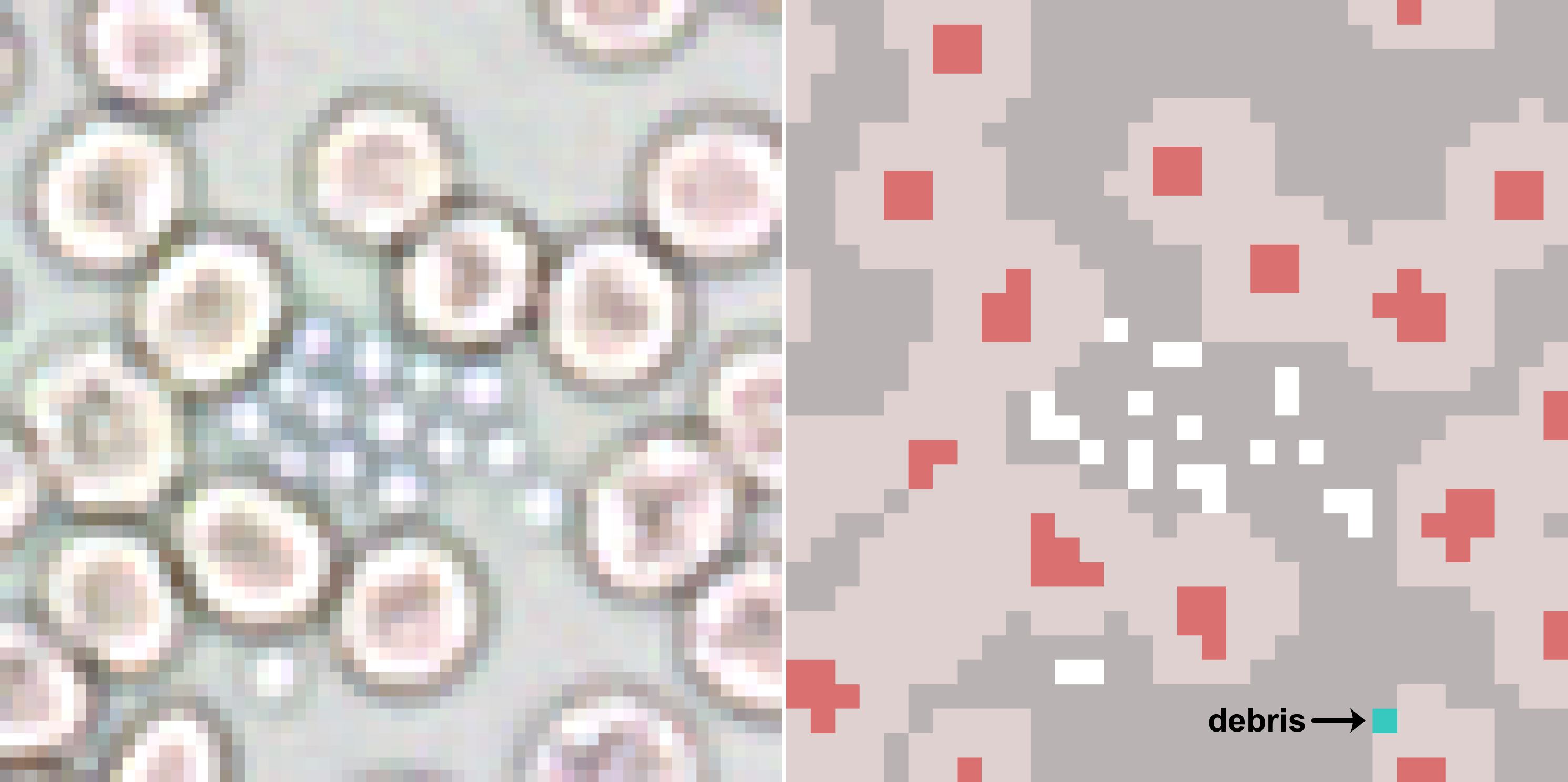}
        \label{fig:u16}
    }
    \subfigure[U-Net-32] 
    {
        \includegraphics[width=0.48\textwidth]{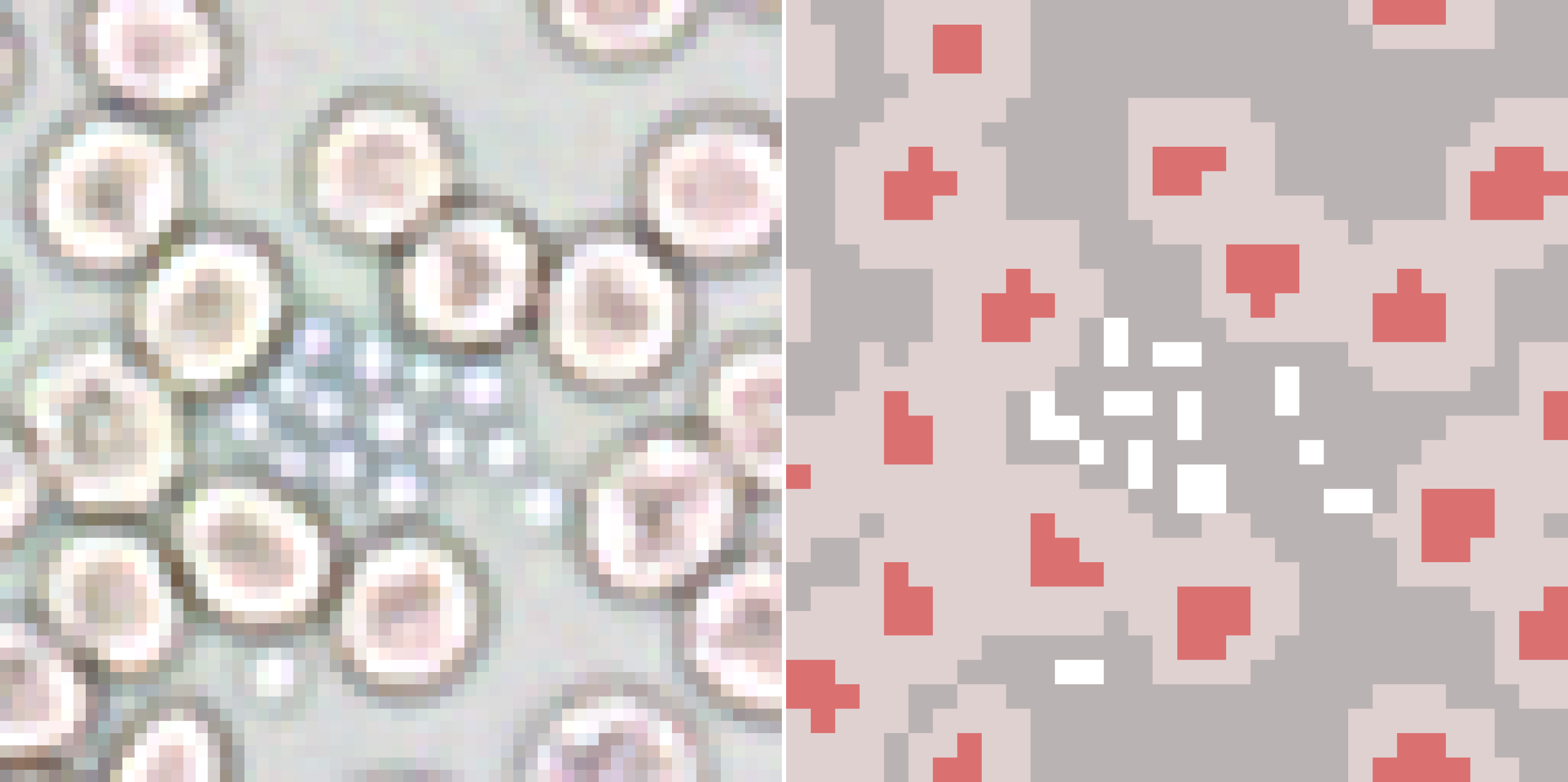}
        \label{fig:u32}
    }
    \subfigure[U-Net-64] 
    {
        \includegraphics[width=0.48\textwidth]{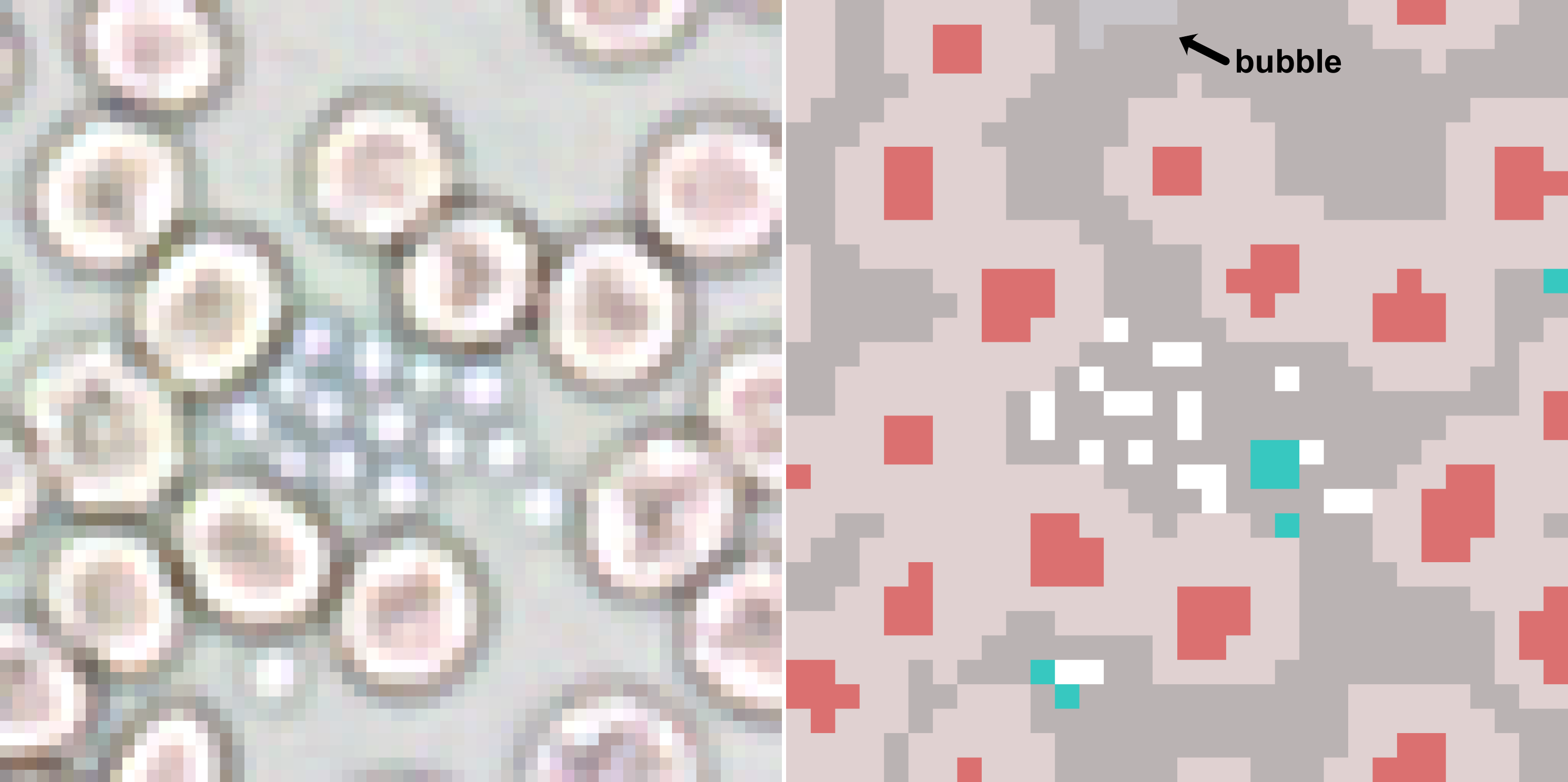}
        \label{fig:u64}
    }
    \subfigure[U-Net-128] 
    {
        \includegraphics[width=0.48\textwidth]{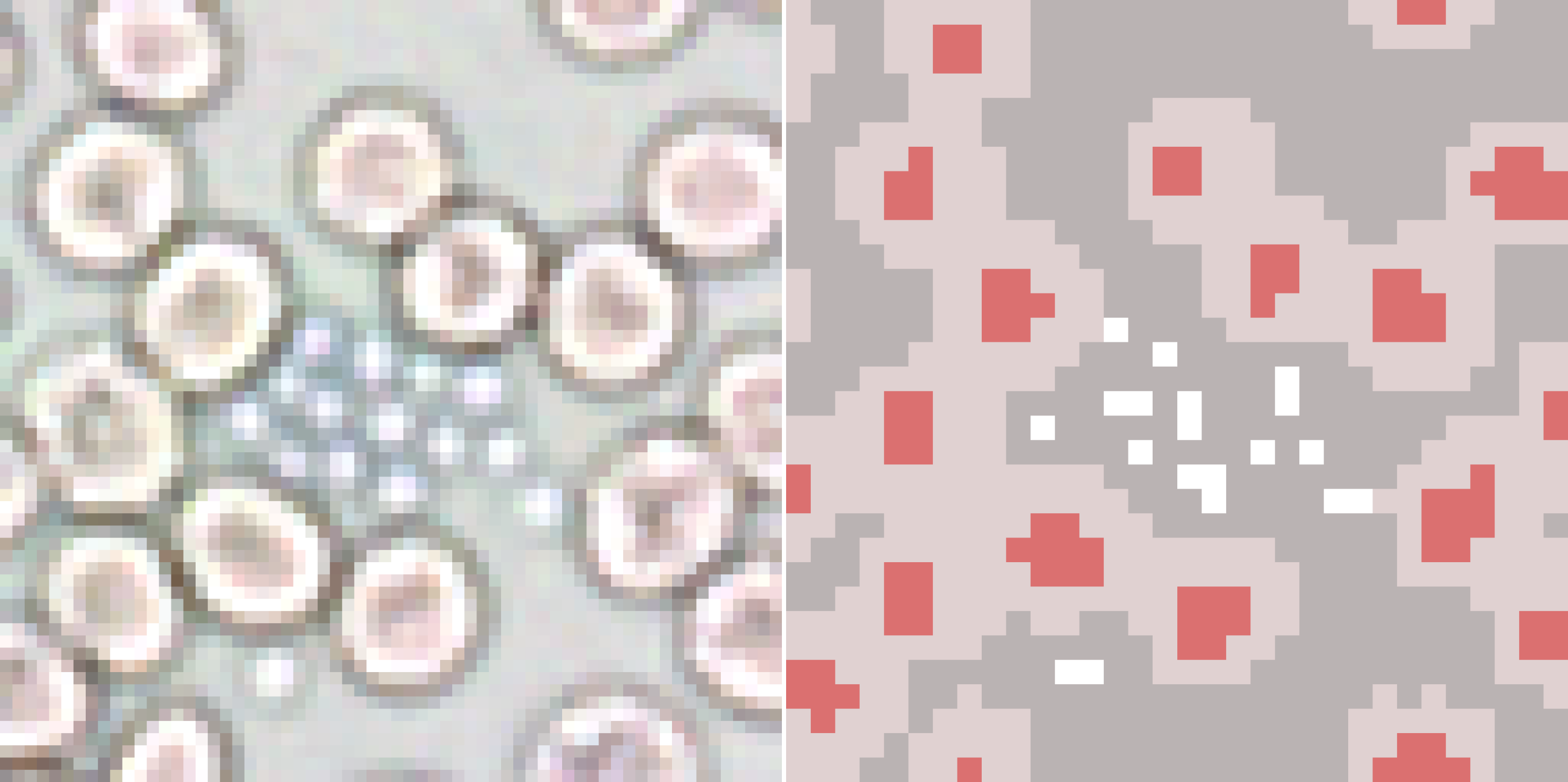}
        \label{fig:u128}
    }
    \caption{High-resolution images and corresponding segmentation masks from different U-Net networks (see Table~\ref{tab:f1_scores} for details). Platelet count for the aggregate is 14.}
    \label{fig:segmentation_mask}
\end{figure*}

\FloatBarrier

\subsection{Comparison of Segmentation Masks}

In this section we compare the segmentation masks produced by several U-Net networks (Fig.~\ref{fig:segmentation_mask}).
Panels (Fig.~\ref{fig:u16},~\ref{fig:u32},~\ref{fig:u64}, and~\ref{fig:u128}) show the results when using 2-stage down- and upsampling blocks for different sized networks.
By comparing the different panels for segmentation masks, we observed variability in the pixels identified as platelets, which has a direct impact on their counts.
In addition, the networks identified pixel as debris (Fig.~\ref{fig:u16}~and~\ref{fig:u64}) which can also influence platelet counts.
We observed a similar result when using the 1-stage down- and upsampling blocks and a single platelet class (Fig.~\ref{fig:u64sl}), where pixels where labeled as white blood cells.
When using the same network with 2 platelet classes (single platelets and platelet aggregates) we obtain very distinctive class identification from the segmentation mask (Fig.~\ref{fig:u64_2_classes}).
Recall from Section~\ref{sect:methods} that a bounding box is positioned over the platelet aggregate to identify peaks from high pixel values. 
In this case, an approach that can be easily achieved from nearby red blood cells and background pixels.

\subsection{Evaluation of Platelet Counting Methods}

Our first approach to evaluate the counting of platelets involved the pixel area method applied to platelet aggregates (\textit{i.e.}, clusters) identified using the density-based spatial clustering as described in Section~\ref{sect:methods}.
This involves the counting of platelets by calculating the number of pixels corresponding to the platelets in an aggregate, followed by normalization by the average number of pixels in a single platelet (defined as 3).
The approach is known to only provide moderate accuracy; in part, due the variability of platelet size and over-lapping platelets and/or blood components.
In Table~\ref{tab:platelet_counts}, we provide the results of this approach, where we compare the standard errors (SE) and coefficient of variations (CV) of the methods with respect to the increasing number of neighboring platelets.
Even with a single platelet, this method results with high variability of results (CV 2.0) and error (SE 0.4).
As we move to 2 platelets, we observed that the results are getting worse (CV 2.7 and SE 0.6).
Errors continue for platelet aggregates of size 3 (CV 2.7 and SE 0.6) and 4 (CV 3.7 and SE 1.4), after which errors start to diverge quickly.
Overall, the pixel area method provided a CV 3.3 and SE 0.5 which would not be acceptable for high-risk applications, such as hematology analytics.

Another common method used for counting objects in images is connected component analysis (CCA).
As mentioned in Section~\ref{sect:introduction}, there are two versions of this approach that is used: 4- and 8-components.
In our case, the 4-component was used to count platelets from segmentation masks.
The segmentation mask from a network is filtered for platelet pixels which is then processed by the CCA algorithm.
A platelet is defined where continuous pixels exists.
If a non-pixel exist in the left-right-up-down direction, the defined platelet ends.

\noindent
For example, Fig.~\ref{fig:u64sl} would result with a count of 10 platelets.
CCA methods often reports accurate counts; in particular, when combined with a neural network (\textit{e.g.}, U-Net).
We report the results of this approach in Table~\ref{tab:platelet_counts}, where CVs and SEs of 0.0 and 0.0, 0.4 and 0.2, 0.5 and 0.2 and 0.8 and 0.5 were obtained for single platelet and aggregates 1, 2, 3, and 4, respectively.
For platelets aggregates greater than 4, we observe both CV and SE to degrade; however, much less than for PAM.
This approach represents one of the state-of-art methods, providing accurate and reliable counts.
However, this approach performs best when platelets are well separated and are not over-lapping.
In addition, this approach is susceptible to noise where small particles may be mistaken for other cells or background, leading to false positives.
For examples, Fig.~\ref{fig:u64sl} identifies several pixels as white blood cells, which is clearly not the case.

Our proposed counting method (PCM) used platelet aggregates (\textit{i.e.}, clusters) identified using DBSCAN as was done for PAM (\textit{vide supra}).
In this case, after localizing each platelet aggregate, individual platelet counts were defined by applying a threshold on values above 0.9, relative to neighboring pixel values.
Recall that platelets have a characteristic feature where central pixels have high intensities.
The results are provided in Table~\ref{tab:platelet_counts}, which performed better than PAM and comparable to CCA.
For a single platelet the PCM resulted with a CV 0.7 and SE 0.3, which is better than PAM but slightly less than the CCA results.
A similar trend was observed for all other platelet aggregate sizes; however, for larger sized platelet aggregates (> 4), the proposed PCM performed slightly better than CCA (CV/SE: 2.3/0.4 vs 2.4/0.9).

\begin{table}[!htb]
\begin{adjustbox}{width=0.48\textwidth}
    \centering
    \begin{threeparttable}
    \caption{Comparison of platelet count estimations and errors using various methods.\tnote{1,2}}
    \begin{tabular}{l l l l}
        \toprule
        \makecell[l]{\textbf{Actual Platelet} \\ \textbf{Count}} 
        & \makecell[l]{\textbf{Pixel Area} \\ \textbf{Method (error)}}
        & \makecell[l]{\textbf{Peak Cluster} \\ \textbf{Method (error)}}
        & \makecell[l]{\textbf{Connected Component} \\ \textbf{Analysis (error)}}
        \\ 
        \midrule
        1 & 2 (1) & 1 (0) & 1 (0) \\
        1 & 2 (1) & 3 (2) & 1 (0) \\
        1 & 3 (2) & 1 (0) & 1 (0) \\
        1 & 3 (2) & 2 (1) & 1 (0) \\
        1 & 4 (3) & 1 (0) & 1 (0) \\
        1 & 4 (3) & 2 (1) & 1 (0) \\
        \midrule
        SE (CV) (1 platelet):
        & 0.4 (2.0)
        & 0.3 (0.7)
        & 0.0 (0.0)
        \\
        \midrule
        2 & 1 (0) & 1 (1) & 2 (0) \\
        2 & 3 (1) & 1 (1) & 3 (1) \\
        2 & 3 (1) & 3 (1) & 2 (0) \\
        2 & 4 (2) & 3 (1) & 4 (2) \\
        2 & 4 (2) & 4 (2) & 2 (0) \\
        2 & 6 (4) & 1 (1) & 2 (0) \\
        2 & 6 (4) & 3 (1) & 2 (0) \\
        2 & 8 (6) & 4 (2) & 2 (0) \\
        2 & 6 (4) & 4 (2) & 3 (1) \\
        \midrule
        SE (CV) (2 platelets):
        & 0.6 (2.7)
        & 0.2 (1.3)
        & 0.2 (0.4)
        \\
        \midrule
        3 & 2 (1) & 1 (2) & 4 (1) \\
        3 & 2 (1) & 2 (1) & 4 (1) \\
        3 & 2 (1) & 4 (1) & 2 (1) \\
        3 & 4 (1) & 2 (1) & 3 (0) \\
        3 & 4 (1) & 4 (1) & 4 (1) \\
        3 & 5 (2) & 3 (0) & 3 (0) \\
        3 & 5 (2) & 4 (1) & 3 (0) \\
        3 & 7 (4) & 1 (2) & 3 (0) \\
        3 & 8 (5) & 2 (1) & 3 (0) \\
        3 & 9 (6) & 3 (0) & 4 (1) \\
        3 & 9 (6) & 2 (1) & 3 (0) \\
        \midrule
        SE (CV) (3 platelets):
        & 0.6 (2.7)
        & 0.2 (1.0)
        & 0.2 (0.5)
        \\
        \midrule
        4 & 2 (2) & 4 (0) & 1 (3) \\
        4 & 4 (0) & 1 (3) & 4 (0) \\
        4 & 5 (1) & 2 (2) & 4 (0) \\
        4 & 8 (4) & 4 (0) & 3 (1) \\
        4 & 10 (6) & 5 (1) & 3 (1) \\
        4 & 13 (9) & 3 (1) & 4 (0) \\
        \midrule
        SE (CV) (4 platelets):
        & 1.4 (3.7)
        & 0.5 (1.2)
        & 0.5 (0.8)
        \\
        \midrule
        5  & 4  (1) & 4 (1) & 3 (2) \\
        6  & 6  (0) & 3 (3) & 5 (1) \\
        6  & 13 (7) & 5 (1) & 6 (0) \\
        6  & 13 (7) & 8 (2) & 6 (0) \\
        8  & 4  (4) & 5 (3) & 4 (4) \\
        9  & 7  (2) & 5 (4) & 3 (6) \\
        11 & 25 (14) & 8 (3) & 11 (0) \\
        16 & 24 (8) & 17 (1) & 10 (6) \\
        \midrule
        SE (CV) (> 4 platelets):
        & 1.6 (5.4)
        & 0.4 (2.3)
        & 0.9 (2.4)
        \\
        \midrule
        SE (CV) (overall):
        & 0.5 (3.3)
        & 0.2 (1.3)
        & 0.2 (0.8)
        \\
        \midrule
        \bottomrule
    \end{tabular}
    $^1$ Error calculated as the absolute difference of the predicted and actual platelet counts. Standard error (SE) and coefficient of variation (CV) computed based on error values for a given method. Standard deviations calculated using sample populations (\textit{i.e.}, $n-1)$.
    $^2$ Actual platelet counts was done by manual inspection. Pixel area and peak cluster methods was evaluated using U-Net-64S2+ network.
    \label{tab:platelet_counts}
    \end{threeparttable}
\end{adjustbox}
\end{table}

\FloatBarrier

\begin{figure}[!htb]
    \centering
    \includegraphics[width=0.48\textwidth]{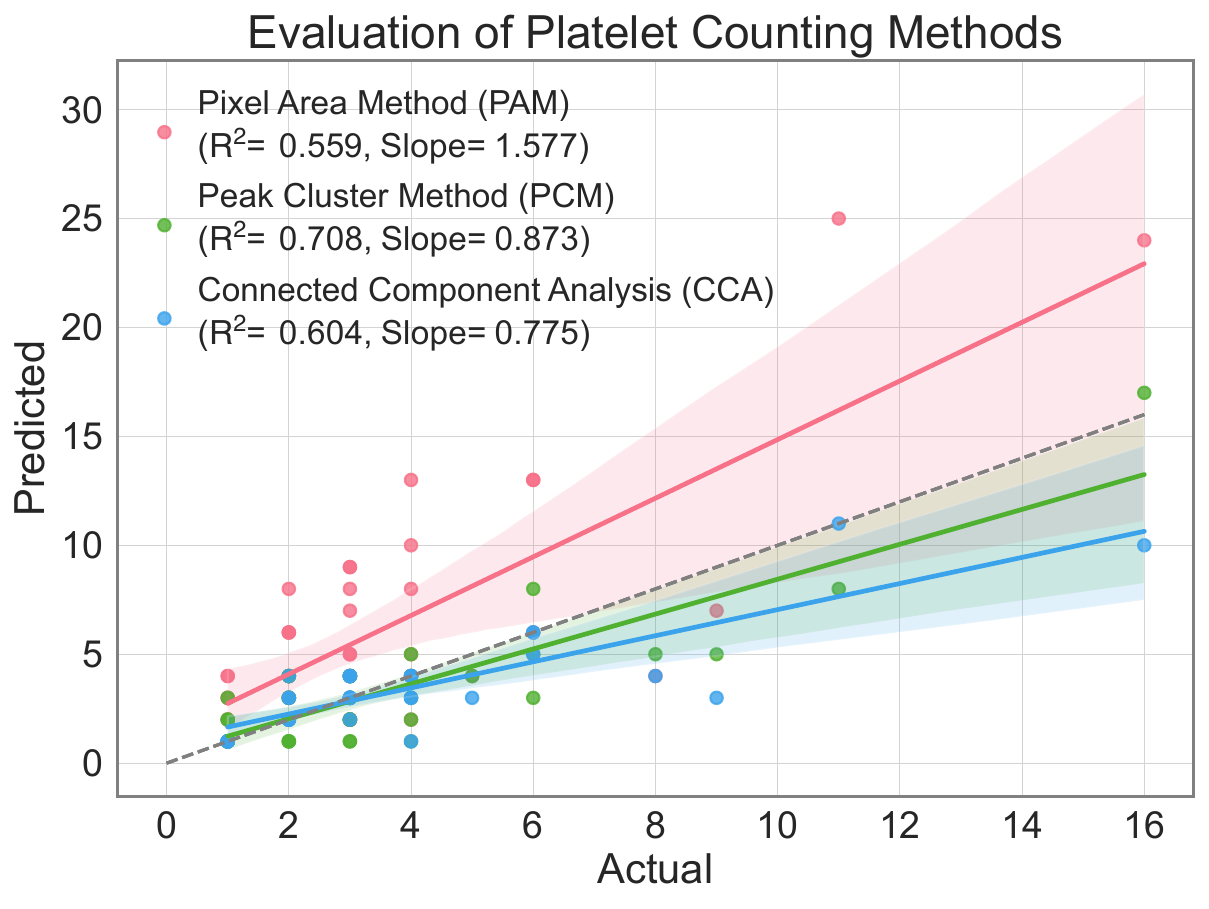}
    \caption{Regression analyses of counting the PAM, PCM, and CCA methods. Regression was performed by forcing a zero y-intercept. Shaded are refers to standard deviations for respective methods.}
    \label{fig:platelet_counting_methods}
\end{figure}

In Fig.~\ref{fig:platelet_counting_methods}, we performed regression analyses for the three different counting methods.
Consistent with the above discussion, PAM perform less than optimal, while both the CCA and PCM provide similar and reasonable results. 
It is noteworthy the all methods struggle to varying degrees with increasing size of platelet aggregates, whereas for smaller sizes (< 6) either CCA or PCM provide accurate predictions and offer viable solutions for counting platelets. 

\FloatBarrier

% CONCLUSION
\section{Conclusions}
\label{sect:conclusions}
In this study we investigated three different platelet counting methods.
We explored U-Net network configurations with different down- and upsampling kernel sizes and added a platelet aggregate class for semantic segmentation.
We demonstrated that the addition of the platelet aggregate class improves F1-scores and the proposed pixel cluster method is a competitive alternative to connected component analysis.
While this study focused on blood components and the counting of platelets, we expect these methods to be applicable to other domains with small objects and their enumeration.

% REFERENCES
\bibliographystyle{IEEEtran}
\bibliography{references}

\end{document}